\documentclass{article}
\usepackage{graphicx}

\begin{document}
\baselineskip 15pt
\begin{center}

{\Large New type of antiferromagnetic polaron and bipolaron in HT$_c$ - superconductors}

Ewelina Hankiewicz, Ryszard Buczko and Zbys{\l}aw Wilamowski

{Institute of Physics, Polish Academy of Sciences, Al. Lotnik\'ow 32/46, PL 02--668 Warsaw Poland\\ email
sabcia@ifpan.edu.pl}

\end{center}
\vspace{.5cm}

Abstract: \vspace{.1cm}

\mbox{}\hfill\parbox{10cm}{The possibility of formation of a new type of polaron based on the quantum
aniferromagnet (AF) model is reported. We take into account exchange interactions between localized $d$-$d$
spins  of the AF, as well as the $p$-$d$ interaction of the AF with $p$-carriers.  The energy minimum  is
found when maximum charge density occurs on every second spin. The formation of such ``comb''-like polarons
results from the damping of quantum fluctuations and the appearance of Van Vleck-like staggered magnetization.
Such polarons tend to form pairs coupled by an AF ``glue''.}\hfill\mbox{}

\vspace{.5cm}

There are some convincing arguments that occurrence of small polarons and bipolarons
is responsible for the formation of a Bose liquid in HT$_c$-superconductors
\cite{alexandrov,Zhang1}. The mechanism of polaron and bipolaron formation, however,
still remains an open question. Theoretical models of this effect invoke the phonons,
excitons, plasmons as well as a magnetic ``glue'' \cite{Mott5,Nature6}.

In this paper we show that the spin exchange interactions alone can lead to the
formation of magnetic polarons. Moreover, a specific type of magnetic polarons, called
by us the ``comb-like'' polarons, is characterized by a strong tendency for pair
formation. For CuO$_2$-based superconductors their binding energy is comparable to
$d$-$d$ coupling, $J_{d-d}$, {\em i.e.}, of the order of 7.5 meV. We solve numerically
the spin Hamiltonian which contains the term of $d$-$d$ coupling between neighboring
$d$-spins within the antiferromagnetic (AF) cluster, $2J_ {d-d} \, ${\mbox{\boldmath
$S_i\cdot S_{j}$}, and the term of $p$-$d$ exchange between the $p$-carrier and the
$d$-spins, $J_{p-d}(i)\,${\mbox{\boldmath $S_i \cdot\sigma$}}. The $p$-$d$ exchange
constants $J_{p-d}(i)$ are assumed to be proportional to the carrier density at the
$i$-th spin. The normalization condition of the carrier wave function guarantees that
the sum of $J_{p-d}(i)$ does not depend on the distribution, but  is equal to a
parameter $N_o\alpha$ which describes the strength of the $p$-$d$ coupling. We examine
various polaron shapes.

 We found ``comb-like'' polarons, where the electron density is distributed on spins from
 only one of Neel sublattices, as the most energetically favorable (see Fig.1(a)).
  The total magnetic moment of such polaron,
$S^*$, is equal to $1/2$ and does not change with an increase of $p$-$d$ coupling. The
net magnetic moment of $d$-spins remains zero but, because of a nonvanishing
correlation between $p$ and $d$-spins, an energy gain due to $p$-$d$ coupling occurs.
The correlators $\langle\sigma | S_i\rangle$, i.e., the polarization of $d$-spins seen
by the carrier spin, is shown in Fig. 1(b).  For a weak $N_o\alpha$ this staggered
magnetization increases linearly with $N_o\alpha$. Then the energy gain originating
from $p$-$d$ coupling, $\Delta E_{1e}$, increases with the square of $N_o\alpha$,
reflecting Van Vleck character of the AF magnetization (see dashed curve in Fig.~2).
The induction of the staggered magnetization is related to stabilization of quantum
fluctuations of the AF cluster. In the inset in Fig. 2 the energy gain, $\Delta
E_{1e}$, is plotted as a function of inverse size of the AF chain. An extrapolation to
large size shows that the energy gain tends to a constant value.
\begin{figure}[btp]
\begin{center}\leavevmode
\includegraphics[width=0.8\linewidth]{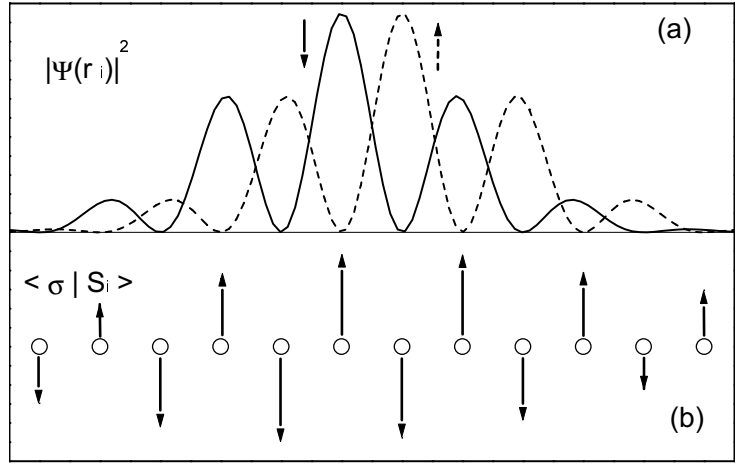}
\caption{ (a) Spatial distribution of carrier density in a ``comb-like'' bipolaron;
(b) Spin polarization of localized $d$-spins, $S_i$, as seen by the carrier $p$-spins,
$\sigma$. The  staggered magnetization caused by a ``comb-like'' monopolaron has a
similar character, but a smaller amplitude.}\end{center}\end{figure}

  The quadratic dependence of  $\Delta
E_{1e}(N_o \alpha)$ shows the possibility of formation of polaron pairs. In Fig. 2 the
energy of the ground singlet, $S^*=0$, and the first excited triplet of the bipolaron
, $S^*=1$, are shown.  The binding energy $E_b$ of such a pair is equal to $\Delta
E_{2e}-2\,\Delta E_{1e}$.  For a weak $p$-$d$ coupling, $E_b$ is equal to 3/4 of the
singlet-triplet energy distance $J_{p-p}$.  The $p$-spins are antiferromagnetically
correlated at the singlet, and ferromagnetically at the triplet state. Thus, one can
conclude that two carrier spins are  effectively coupled by an indirect Heisenberg
exchange, $J_{p-p}\;${\boldmath $\sigma \cdot\sigma$}, where the AF cluster plays a
role of a ``glue''.

In CuO$_2$-based superconductors, however, $ N_o \alpha\,\approx 10\,J_{d-d}$. As a
consequence, the linear response approximation fails. The energy gains, the binding
energy, and the $p$-$p$ exchange all saturate. In the inset in Fig. 2 the saturation
value of $J_{p-p}$ is plotted. For a small cluster size it is of the order of
$J_{d-d}$, {\em i.e.}, it is comparable or greater than the binding energy caused by
the phononic effect. With an increase of the cluster size the parameter $J_{p-p}$
tends to zero. This allows us to conclude that a short AF coherence range (observed in
HT$_c$--materials) is the necessary condition for the existence of an AF ``glue''.
\begin{figure}[btp]
\begin{center}\leavevmode
\includegraphics[width=1.05\linewidth]{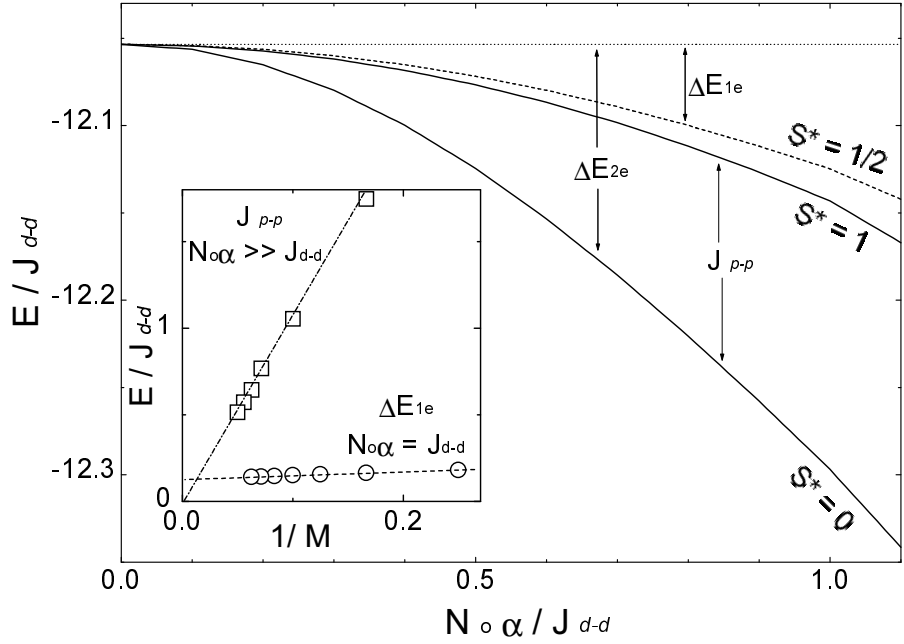}
\caption{Energy levels of the polaron (dashed) and of the  bipolaron (solid curves) as
a function of $p$-$d$ coupling $N_o\alpha$. The inset shows the dependence of the
energy gain of the polaron, $\Delta E_{1e}$, for a small $N_o\alpha$, and the
effective indirect exchange, $J_{p-p}$, between carrier spins in the bipolaron for
very large $N_o\alpha$ as a function of inverse cluster size.}
\end{center}\end{figure}

Work supported by KBN grant 2 P03B 0007 16.

\end{document}